Quantifying Strain and its Effect on Charge Transport in Ge/Si Core/Shell Nanowires

*Aswathi K. Sivan, Nicolas Forrer, Aakash Shandilya, Yang Liu, Janica Böhler, Alexander Vogel, Arianna Nigro, Pierre Chevalier Kwon, Artemii Efimov, Ilya Golokolenov, Gerard Gadea, Riccardo Rurali, Andreas Baumgartner, Dominik M. Zumbühl, and Ilaria Zardo\**

A. K. Sivan, N. Forrer, A. Shandilya, Y. Liu, J. Böhler, A. Vogel, A. Nigro, P. Chevalier Kwon, A. Efimov, I. Golokolenov, A. Baumgartner, D. M. Zumbühl, and I. Zardo
Department of Physics, University of Basel, Basel, 4056, Switzerland
E-mail: Ilaria.zardo@unibas.ch

A. Vogel, G. Gadea, A. Baumgartner, and I. Zardo
Swiss Nanoscience Institute, University of Basel, Basel, 4056, Switzerland

R. Rurali
Institut de Ciència de Materials de Barcelona, ICMAB–CSIC, Campus UAB, Bellaterra, 08193, Spain

Funding: This work was supported as a part of NCCR SPIN, a National Centre of Competence in Research, funded by the Swiss National Science Foundation (grant number 225153), the Georg H. Endress Foundation, Nachwuchsförderungs der Universität Basel, the Severo Ochoa Centres of Excellence Program under grant CEX2023-001263-S, and the Agencia Estatal de Investigación under grant PID2024-162811NB-I00. We acknowledge the support by the EU's H2020 Marie Skłodowska-Curie Actions (MSCA) cofund Quantum Science and Technologies at the European Campus (QUSTEC) grant no. 847471, and the UpQuantVal InterReg. We thank the Centro de Supercomputación de Galicia (CESGA) for the use of their computational resources.

Strain engineering in semiconductor nanostructures offers a promising route to optimize electronic and optical properties for advanced quantum technologies. This study explores the relationship between core and shell thicknesses and strain distribution in Ge/Si core/shell nanowires, targeting their application as hosts for spin qubits. Nanowires were synthesized using an Au-catalysed chemical vapor deposition technique, achieving control over core and shell dimensions. High-resolution transmission electron microscopy and elemental mapping confirmed structural integrity, while Geometric Phase Analysis and Raman spectroscopy provided both qualitative and quantitative insights into strain variations driven by core and shell dimensions. Furthermore, polarization resolved µ-Raman measurements allowed us to quantify the longitudinal and transverse phonon mode splitting as a function of strain in the Ge core. The electronic transport properties were investigated by hole mobility measurements. Finally, we observed a record high hole mobility of 25,400 cm² V⁻¹ s⁻¹, underscoring the potential of our core/shell nanowire structures for the realization of high-fidelity spin qubits. Our findings highlight the critical role of geometry in strain tuning and provide valuable design guidelines for optimizing Ge/Si nanowires in scalable quantum device architectures.

## 1. Introduction

Ge/Si heterostructures have been extensively studied over the past decade due to their seamless integration with existing Si electronic platforms, making them an attractive system for advanced device designs.[1,2] In particular, since their theoretical prediction in the late nineties,[3] using spin states for quantum computing has gained significant traction and in this regard, Si/Ge heterostructures are proposed to be promising host systems for spin qubits.[3–7] Hole qubits in Ge have several advantages such as reduced hyperfine interaction due to low natural abundance of spinful nuclei, which helps to reduce qubit decoherence.[8] The strong spin-orbit coupling in Ge/Si structures[7,9] also allows for fast manipulation of spins through all electric means.[10–12] Several Ge/Si architectures are being explored in this regard, such as Ge/Si hut wires,[13] planar heterostructures,[14] and core/shell nanowires (CS NWs).[15–21] In Ge/Si heterostructures, due to the band alignment at the interface, the holes are confined in the Ge, allowing for the possibility of creating hole spin qubits.[22]

In the case of Ge/Si CS NWs, the small diameter of Ge/Si NWs enables their use in superconducting devices operating in the few-channel Josephson junction regime[21,23] and the strong confinement due to the one-dimensional geometry combined with the large band edge offset between the Ge core and the Si shell give rise to a distinct form of spin–orbit interaction known as direct Rashba spin–orbit interaction (DRSOI). [24]  The DRSOI in NW structures is electrically tuneable, a highly desirable feature for quantum computing applications.[25] One of the important parameters for the design of hole-qubits in Ge is the hole mobility ($\mu_h$). The $\mu_h$ in Ge can be controlled by strain and confinement in the Ge/Si heterostructures as strain engineering influences the band offset as well as the effective mass of the holes.[26,27] Indeed, strain is inherently present in Ge/Si heterostructures due to the lattice mismatch between Ge (5.657 Å) and Si (5.431 Å).[28] As a result, in Ge/Si CS NWs, the Ge core experiences compressive strain, while the Si shell is under tensile strain.[29,30] This strain significantly modifies the optical and electronic properties of the NW system.[31,32] It is, therefore, crucial to quantify the strain in order to understand and optimize the charge carrier behaviour, including the $\mu_h$ in Ge/Si CS NWs. One effective strategy to tune the strain is by varying the shell-to-core radius ratio, which directly controls the degree of strain transfer between the shell and the core. However, while strain is an important parameter to tune the material properties, the resulting increase in elastic energy can also lead to the formation of defects through which the material relaxes strain.[33] In this regard, the NW geometry offers a significant advantage, as it allows for elastic strain relaxation through the sidewalls,[34] providing more flexibility in efficient strain engineering. On the other hand, strain engineering is only possible with a reliable and accurate quantification of the strain level in the heterostructures, in combination with the assessment of its impact on their electronic properties.[35]

For the quantification of strain in heterostructures, micro($\mu$)-Raman spectroscopy and high-resolution transmission electron microscopy (HR-TEM) are among the most well-established experimental techniques. $\mu$-Raman spectroscopy serves as a powerful, non-destructive optical technique to probe strain in nanostructures by tracking shifts in phonon frequencies.[36] In strained materials, the phonon energies are blue (red) shifted depending on the compressive (tensile) nature of the strain as compared to their unstrained counterpart. This makes $\mu$-Raman spectroscopy particularly well-suited for quantifying strain in individual Ge/Si CS NWs. Moreover, Raman spectroscopy provides comprehensive insight into the lattice dynamics of the material, which is crucial for device design, as phonons play an essential role in determining the thermal properties of the system. In contrast, HR-TEM allows for the measurement and mapping of strain at the nanoscale using a technique called Geometric Phase Analysis (GPA). GPA is a digital image-processing technique that quantifies local strain and displacement fields in high-resolution microscopy images by extracting and comparing the phase of periodic lattice fringes.

A high carrier mobility is essential for the realization of high-quality qubits. In conventional NW systems, strong disorder and surface scattering can also generate low-frequency charge noise, [35] which in turn leads to qubit decoherence, remaining a major bottleneck in spin-qubit research and technology.[37] Carrier mobility is a fundamental parameter that reflects the crystalline quality and electronic transport properties of NWs, quantifying how efficiently charge carriers respond to an external electric field, [35] characterized through field-effect transport measurements. [38]

In this paper, we present a systematic study of strain engineering in Ge/Si CS NWs with varying core and shell dimensions. Using μ-Raman spectroscopy on individual NWs, we quantitatively analyse the compressive strain in the Ge core as a function of the Si shell thickness. The change in the Raman lineshape from a Lorentzian in bare Ge NWs to a Fano lineshape in the CS NWs indicates the accumulation of hole gas in the Ge.[39] Additionally, we investigate how variations in core diameter influence the strain for a fixed shell thickness. Our results show that the strain in the Ge core increases with increasing Si shell thickness. Furthermore, we use GPA to measure the relative strain in the Si shell with respect to the Ge core. Combined with the Raman spectroscopy, these complementary experimental techniques provide both quantitative and qualitative insights into strain engineering in Ge/Si CS NWs of varying dimensions. We then extend our Raman studies to probe the splitting between the longitudinal optical (LO) and transverse optical (TO) phonon modes of Ge core under strain. Our experimental findings, complemented by *ab initio* density functional perturbation theory (DFPT) calculations, enable a detailed quantification of the energy splitting between the LO and TO modes. Finally, we investigate the transport properties of Ge/Si CS NWs. We demonstrate an average field-effect mobility of 8,000 cm² V⁻¹ s⁻¹, reaching a record value of 25,400 cm² V⁻¹ s⁻¹ in the best-performing device.

## 2. Results and Discussion

2.1. Growth and structural characterization of the CS NWs:

The Ge NWs used in this study were all grown using the vapor-liquid-solid (VLS) approach under a gold-catalysed reaction and for the Ge/Si CS NWs, an uncatalyzed plasma enhanced deposition was added for the Si shell. The Si shell thicknesses were controlled by varying the growth times, keeping the other deposition parameters fixed.

In order to study the strain level as a function of core and shell dimensions, two different series of Ge/Si CS NWs samples were prepared: i) a series with a fixed Au colloids size (15 nm) for the Ge cores growth and a varying Si shells thickness (growth times: 2, 3, and 7 minutes) and; ii) a series with varying Ge cores dimension (Au colloids of 5, 10, 15, and 20 nm) keeping fixed the Si shell thickness (growth time: 2 minutes). As reference, Ge NWs were synthesized without Si shells using Au colloids of 5, 10, 15, and 20 nm. In **Table 1**, we summarize the samples used in this study. The diameters of the Ge core samples will be nominally referred to the size of the Au colloid, i.e., 5 nm, 10 nm, 15 nm, and 20 nm.

**Table 1.** Summary of the samples synthesized using VLS growth with commercial Au colloids of varying sizes. The table includes Ge/Si CS NWs with varying Si shell growth times and Ge NWs without Si shells.

| Sample | Au Colloid Size [nm] | Si shell growth time [minutes] |
|---|---|---|
| Ge NWs | 5, 10, 15, 20 | 0 |
| Ge/Si CS NWs | 15 | 2, 3, 7, 15 |
| Ge/Si CS NWs | 5, 10, 15, 20 | 3 |

**Figure 1** (a)-(c) shows HR-TEM images of cross-sections of Ge/Si CS NWs for the Ge core grown with 15 nm Au colloid, with each panel including a 5 nm scale bar and an inset of the selected area electron diffraction (SAED) pattern collected on each cross-section. In Figure 1 (a), the Ge NW with no Si shell consists of a uniform, single-crystalline Ge core (27.5 ± 0.4 nm diameter) with sharp lattice fringes and no detectable amorphous overlayer. A thin oxide layer (1.3 ± 0.6 nm) uniformly surrounds the wire. After 3 min of Si deposition (b), a slightly asymmetric Si shell (with thickness from 4.2 ± 0.8 nm to 5.4 ± 0.8 nm) covers the Ge core (37.3 ± 0.8 nm diameter), preserving clear core–shell contrast and coherent interfaces; the SAED pattern confirms the core's crystallinity. Extending growth to 7 min (c) yields a thicker and asymmetric Si shell (with thickness from 9.7 ± 0.8 nm to 13.8 ± 0.8 nm) that remains continuous around the core (29.3 ± 0.8 nm diameter). These micrographs show that the plasma deposited shell indeed leads to a crystalline Si shell around the Ge core. For both designs, a 1.8 ± 0.8 nm thick layer of native $SiO_2$ can be observed. The high crystallinity of the wires is further supported by the SAED pattern in the insets. Line profiles obtained from Electron Dispersive X-ray (EDX) analysis of the pure Ge NW and Ge/Si CS NWs are displayed in Figure 1 (d), showing the elemental distribution of Ge in red, Si in blue, O in light blue, and C in orange. There are several interesting things to be observed from this elemental analysis. In the centre of all three NWs we observe a pure Ge core, with minimal background signals of the other elements. At the edges of the Ge NW (Figure 1 (a) and top panel of Figure 1 (d)) we note a rise in the O signal, indicating the formation of a thin oxide layer at the NW surface. In the Ge/Si CS NWs, displayed in Figures 1 (b, c) and middle and bottom panels of Figure 1 (d), we observe a sharp rise in the Si signal at the edges of the Ge core, without increase in the O signal at the Ge/Si interface. The thickness of the Si layer increases, as expected, with increasing shell growth time. Finally, we observe an increase in the O signal at the edges of the Si shells, in line with the formation of a native oxide layer. These results demonstrate precise, time-dependent control over the Si shell thickness and composition on Au-seeded Ge NWs. However, an uneven distribution of the Si shell around the Ge core is also apparent. We attribute the origin of this uneven distribution to shadowing effects due to the high NWs density.[40] Overall, the shell deposition observed fits well with what we expect, i.e., longer growth times lead to thicker shells.

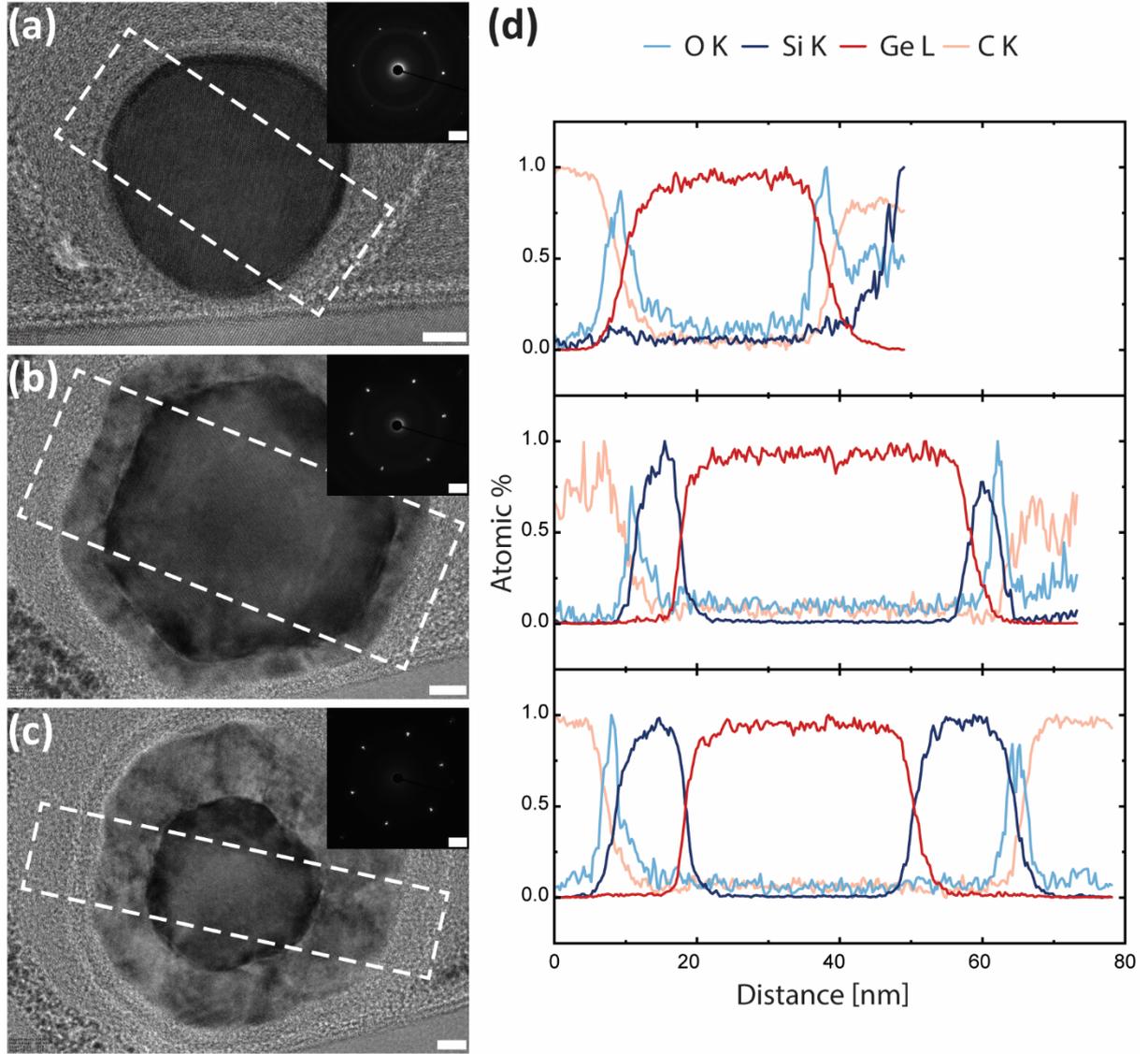

**Figure 1.** (a) – (c) HR-TEM images of NW cross-sections for a (a) Ge NW grown from a 15 nm Au colloid, (b,c) Ge/Si CS NW grown from a 15 nm Au colloid with a shell thickness of 6.5 ± 2 nm (b), and 11 ± 2 nm (c), respectively. The insets show a SAED pattern obtained on each cross-section. (d) EDX line profiles of the O-K, Si-K, Ge-L, and C-K edges obtained on the cross-sections, as indicated in panels (a) – (c). The scale bar for the HR-TEM images and for the SAED patterns is 5 nm and 2 nm$^{-1}$ in each panel, respectively.

We obtain the strain in the Si shell relative to the Ge core using GPA on the HR-TEM images. In **Figure 2**, qualitative maps of the relative strain are displayed. From these maps, we see a uniform tensile relative strain in the Si shell of both 6.5 ± 2 nm and 11 ± 2 nm thickness. To quantify the strain in the Si shell with respect to the axial and transverse direction of the wire, we extract a section of the Ge core/Si shell area and align it with the image frame. Using this approach, we calculate an axial relative strain $\varepsilon_{xx} = -3.4\%$ and transverse relative strain $\varepsilon_{yy} = -2.9\%$ in the 6.5 ± 2 nm shell, and $\varepsilon_{xx} = -4.7\%$ and $\varepsilon_{yy} = -4.1\%$ in the 11 ± 2 nm one. The relative strain extracted from GPA analysis as a function of γ-parameter, where γ is a dimensionless geometric parameter that describes the ratio of the core radius to the total radius of the core/shell NW, is displayed in Figure 2 (c). As demonstrated in literature,[41] large errors in GPA-based strain measurements from HR-TEM data are strongly dependent on image resolution, particularly for thin lamellas. Consistent with this, our previous results[42] show that higher-resolution imaging leads to a significant reduction in strain quantification error. The γ-parameter provides a measure of how much of the NW's cross-section is occupied by the core.

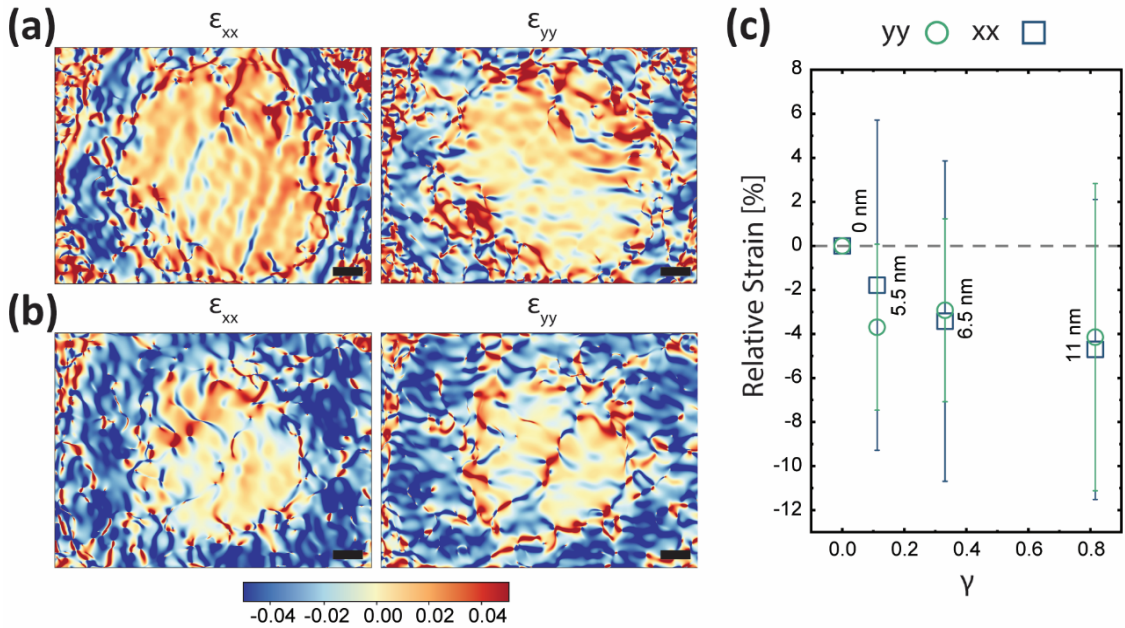

**Figure 2.** Relative strain GPA maps obtained via GPA of the cross-sectional HR-TEM for Ge/Si CS NW grown from a 15 nm Au colloid with a shell thickness of 6.5 ± 2 nm (a), and 11 ± 2 nm (b), respectively. Rotated relative strain of the shell compared to the core for different shell thickness as presented in Table 1. (c) Relative strain of the Si shell extracted from GPA analysis as a function of γ-parameter, with γ the ratio of the core radius to the total radius of the core/shell NW. Here the x- and y-directions were aligned with the radial and transversal axis of the wire, respectively.

2.2. Micro-Raman Spectroscopy:

We performed μ-Raman measurements on individual CS NWs transferred on TEM grids, which were later used for microscopy measurements and GPA analysis.

To investigate the effect of Si shell thickness on the strain induced in the Ge core, we fixed the core diameter and varied the Si shell growth times, i.e. Si shell thickness. The Ge core was synthesized using 15 nm diameter Au colloids. The results of Raman measurements are presented in **Figure 3**. For the sake of comparison, we also measured 15 nm Ge NWs without any Si shell: these measurements were used as a reference for calculating strain induced Raman shifts and corresponding strain values.

In Figure 3(a), we present Raman spectra taken from CS NWs of varying Si shell thicknesses, the black dotted line is a guide to the eye to indicate the frequency of the degenerate TO/LO phonon mode at the Γ point of Ge NW without the Si shell. It is evident that as the Si shell thickness increases, the Ge TO/LO peak position shifts to higher wavenumbers, indicating increasing compressive strain in the Ge core. To address potential non-uniformities and inhomogeneities in core diameters during growth, we conducted µ-Raman measurements on several individual NWs of the same as-grown sample. To be consistent, data were collected from the centre of the NWs to account for tapering during the growth. The detailed results of these measurements are provided in the Supporting Information (SI). The Raman spectrum from bare Ge NW was fitted with a Lorentzian curve, while the data of the CS NWs were fitted with a Fano line shape[43] given by:

$$I(\omega) = A \frac{(q+\epsilon)^2}{1+\epsilon^2}, \epsilon = \frac{\omega - \omega_0}{\Gamma} \qquad (1)$$

where q is the Fano asymmetry parameter, Γ is the linewidth parameter and $\omega_0$ is the phonon frequency.[43,44] The presence of Fano asymmetry is expected and is attributed to the accumulation of holes in Ge core in the Ge/Si CS NWs.[45]

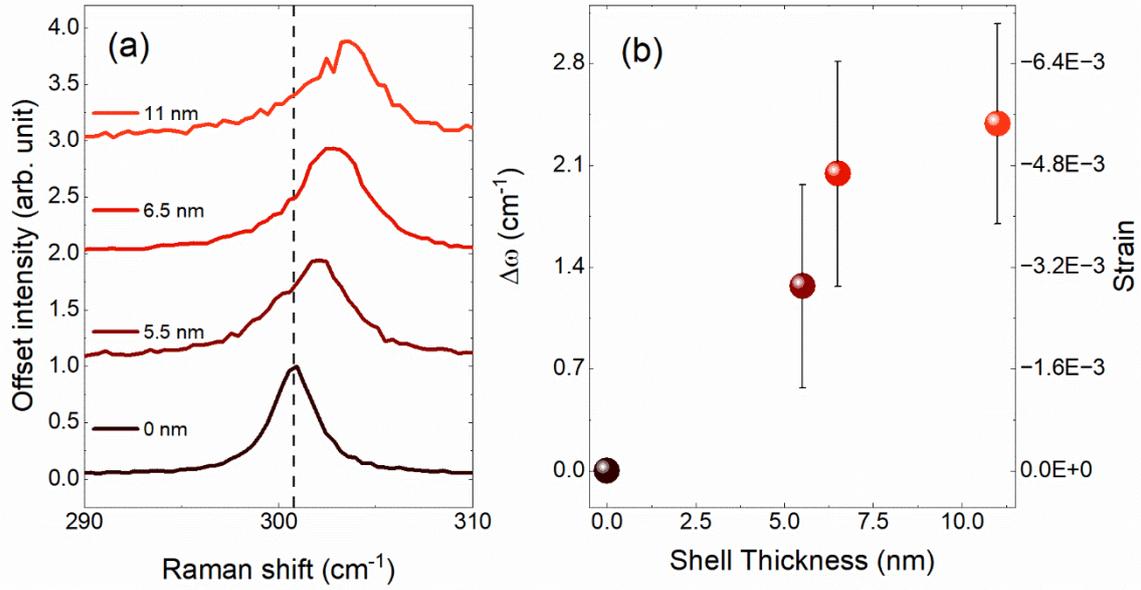

**Figure 3.** (a) Waterfall plot of μ-Raman spectra of Ge/Si CS NWs with varying Si shell growth thickness for a fixed colloids size for the Ge core (15 nm). The spectra are normalized to their maximum intensity. The dotted line is guide to eye, pointing the frequency of the TO/LO phonon mode of Ge NWs without Si shell; (b) Raman peak shift (Δω, left y-axis) and corresponding absolute strain values (right y-axis) calculated from the Raman peak shifts as a function of Si shell thickness for Ge/Si CS NWs grown with 15 nm Au colloid.

We measured the average peak position of the Ge TO/LO phonon mode as a function of Si shell thickness from multiple individual NWs per sample. From the Raman shift of the Ge phonon mode, we could quantify the strain experienced by the Ge core due to the Si shell, using the equation[46]:

$$\omega(\epsilon) = \omega_0 + b\epsilon \tag{2}$$

with

$$\epsilon = \frac{(\omega - \omega_0)}{b} = \frac{\Delta\omega}{b} \tag{3}$$

where ω is the Raman shift of the Ge phonon mode in the CS NW, $\omega_0$ is the relaxed Raman frequency, b is the phonon strain shift coefficient, and $\epsilon$ is the strain. There are different values of strain shift coefficients in literature; in this work, we use the value reported by Pezzoli et al. [46] and equal to $-440 \pm 8$ cm$^{-1}$. The negative strain shift coefficient means that tensile strain leads to a redshift (decrease in Raman frequency), while compressive strain causes a blueshift (increase in Raman frequency). For $\omega_0$ we used the frequency of the Raman peak of the reference Ge NW without the Si shell. The Raman shifts ($\Delta\omega = \omega - \omega_0$) as a function of the shell thickness, along with the corresponding strain level are plotted in Figure 3(b). Specifically, the left y axis of Figure 3(b) shows the Raman shift $\Delta\omega$ obtained by subtracting the phonon frequency of the bare Ge NWs of the same Au colloid size from that of Ge phonon mode in CS NWs of different Si shell thicknesses. The shift of the Ge phonon mode increases with increasing shell thickness, indicating a progressive increase of strain in the Ge core with increasing shell thickness. The right y axis in panel (b) presents the corresponding strain evolution, revealing an increase in compressive strain in the Ge core as the Si shell thickness increases. This trend is expected due to the lattice mismatch between Ge and Si, where the growing Si shell imposes compressive stress on the Ge core. However, the strain appears to saturate for thicker shells, suggesting the onset of strain relaxation mechanisms such as defect formation or surface/interface-driven strain redistribution. This saturation behaviour is characteristic of CS NWs,[47] where strain initially builds up elastically due to lattice mismatch but becomes limited by relaxation processes at higher shell thicknesses. We observe a maximum value of -0.005 ± 0.001 of strain for CS NWs with 11 ± 2 nm Si shell for the CS NWs grown with 15 nm Au catalyst. Additionally, Raman measurements from a sample with a 16 ± 3 nm Si shell on a 15 nm Ge core (see SI) show a broad, red-shifted peak, characteristic of defect-induced strain relaxation. Previously, strain analysis in Ge/Si CS NWs have been carried out using X-Ray diffraction and HR-TEM measurements.[48,49] In particular, in the work done by Goldthorpe and co-authors,[48] using TEM and synchrotron X-ray diffraction, they observed that dislocation formation and stress-driven surface roughening facilitate the relaxation of the strain as the thickness of the Si shell increases for a fixed Ge core. They also observed that the strain also increases as diameter of core decreases for a given shell thickness. In their CS NWs with a Ge core of 26 nm and a shell of 6 nm thickness, they measured a compressive strain of 0.77 %. When the thickness of the shell was increased to 17 nm, they observed a strain of 0.47 %. In our case, we observe that the strain increases as we increased the Si shell to 11 nm (0.54 %) and then strain relaxes through defect formation for thicker shells.

As mentioned above, the spectra collected from CS NWs were fitted with a Fano line shape, consistent with a system where accumulation of carriers occurs. In **Table 2**, we summarize the q values (see Eq.1) obtained through the Fano fits of our CS NWs with different Si shell thicknesses. The error is the standard deviation from the fitting of measurements on multiple wires per CS design. A lower q value indicates stronger Fano asymmetry, reflecting enhanced coupling to the hole continuum. We observe that the average q values for CS NWs with 5.5 ± 1 nm and 6.5 ± 2 nm Si shell thickness remain similar within the margin of error. However, the CS NWs with a 11 ± 2 nm Si shell exhibit a q value of approximately –21, which is notably lower. We compared this value to the Fano asymmetry parameters reported in literature. In a study done by Fukata *et al.*, [45] q values between -30 and -10 where observed for i-Ge/p-Si CS NWs where the hole gas concentration was controlled by changing B-doping concentration in the Si shell. In their work, by doping the Si-shell, the holes flow from the p-Si to the i-Ge region due to the band offset [50] and they estimated a hole density in the Ge to be $10^{17}$–$10^{18}$ cm$^{-3}$. In our work, a q value of -21 suggests that even without intentional doping, thick Si shell may be accumulating holes in the Ge core causing enhanced Fano coupling, mimicking the effects of p-type doping. In Zhang *et al.*, [51] hole gas accumulation in the i-Ge shell layers of p-Si/i-Ge CS NWs was demonstrated using Raman spectroscopy. Their |q| values decreased with increased p-doping of the Si shell due to increased hole accumulation in the Ge. They observed q values between -25 and -5 for varying doping conditions and roughly estimated the hole gas density in the i-Ge shell layers of the p-Si/i-Ge core–shell NWs in the order of $10^{18}$ cm$^{-3}$. A larger value of q was found in work done by Zhang *et al.* [52], comparable to what we observed for our 5.5 ± 1 nm and 6.5 ± 2 nm Si shells.

**Table 2**. The Fano asymmetry parameter, q, obtained from the fit of the spectra collected from Ge/Si CS NWs with 15 nm Ge core and different Si shell thicknesses compared to q values from literature in different systems.

| Si Shell thickness [nm] (this work) | q value (this work) | q value (literature) |
|---|---|---|
| 5.5 ± 1 nm | -41.7 ±11.3 | (~ -45 to -20)[52] – q becomes more negative with increasing Ge shell thickness in p-Si/i-Ge CS NWs. |
| 6.5 ± 2 nm | -43.1 ± 11.0 | (~ -45 to -20)[52] - q becomes more negative with increasing Ge shell thickness in p-Si/i-Ge CS NWs. |

| 11 ± 2 nm | -21.1 ± 7.1 | (~ -30 to -10)[45] – estimated hole density in Ge - $10^{17}$–$10^{18}$ cm$^{-3}$ |
| | | (~ -25 to -5)[51] – estimated hole density in Ge - $10^{18}$ cm$^{-3}$ |

To further understand the influence of different core-diameters on strain, we fixed the Si shell growth times to 2 minutes (i.e. 5.5 ± 1 nm thickness) and varied the core diameters (sample details are in row 2 of Table 1). **Figure 4** presents the Raman frequency and corresponding strain evolution in Ge/Si CS NWs as a function of core diameter. In Figure 4(a), we compare the Raman peak frequency of bare Ge NWs (empty circles) with that of Ge/Si CS NWs (spheres). The values are averaged over measurements on multiple individual NWs and the error bars correspond to the standard deviation. The data show a systematic decrease in Raman frequency with smaller colloid sizes for bare Ge NWs, suggesting an influence of NW diameter on phonon properties. This trend could be attributed to a combination of phonon confinement effects and strain variations induced by changes in NW diameter. A clear upshift in the Raman frequency is observed for the CS NWs compared to their bare counterparts, which is attributed to the compressive strain imposed by the Si shell due to the lattice mismatch between Ge and Si. However, no clear trend can be observed with changing the colloids size (i.e. NWs' diameter). Figure 4(b) shows the corresponding extracted strain values in the Ge core, revealing compressive strain across all core diameters. In other words, the strain does not exhibit a clear monotonic dependence on core size for the 5.5 ± 1 nm Si shell thickness within the investigated core dimensions.

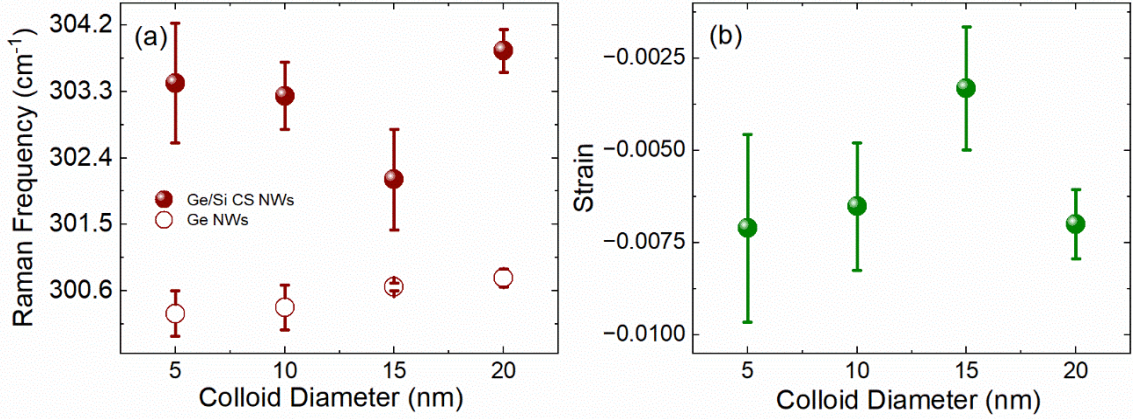

Figure 4 (a) Raman peak frequency of CS NWs with different core diameters and a Si shell thickness of 5.5 ± 1 nm (solid spheres), and Ge NWs with different diameters (empty circles). (b) Strain values calculated from the Raman shift with respect to the bare NWs.

*Understanding the LO-TO splitting in strained Ge/Si CS NWs:*

The presence of strain creates anisotropic modification in the crystal lattice and this alters the phonon dispersion, particularly by lifting the LO and TO degeneracy at the Γ-point in Ge.[53,54] Under compressive strain, the symmetry of the cubic crystal structure is broken and this causes the LO and TO phonons to respond differently, leading to a measurable frequency splitting in Raman spectra. We performed polarization resolved Raman spectroscopy on individual CS NWs to quantify the LO-TO splitting. We use the conventional Porto notation of the form $k_i(\varepsilon_i, \varepsilon_s)k_s$ to indicate the polarization configuration of our Raman measurements, where $k_i$ and $k_s$ are the direction of propagation of incident and scattered photon, respectively, while $\varepsilon_i$ and $\varepsilon_s$ are the direction of polarization of incident and scattered photon, respectively. In this work, since we use backscattering geometry, we assume the incident and scattered photon wave vectors to be antiparallel and parallel to the x axis. Therefore, the polarization vectors lie in the plane perpendicular to the directions of propagation, i.e., the yz plane. In our experiments, we consider the NW growth axis to lie along the z axis. Measurements were done in two polarization configurations: $\bar{x}(z,z)x$ scattering configuration with the polarization of the light parallel to the NW axis, and $\bar{x}(y,y)x$ with the polarization perpendicular to the NW growth axis.

**Figure 5**(a) shows the normalized Raman spectra measured in the $\bar{x}(z,z)x$ (blue) and $\bar{x}(y,y)x$ (pink) polarization configurations for CS NW with a 11 ± 2 nm Si shell grown on a 15 nm Au colloid. A clear shift in phonon frequencies is observed between the two polarization directions in this representative measurement. To better understand these observations, we performed *ab initio* DFPT calculations for compressively strained Ge along the (111) direction. The strain values used in the calculations correspond to those experimentally extracted from Raman measurements, as shown in Figure 3(b). The results are displayed in Figure 5(b), where we present the calculated Raman spectra in 2 different polarization configurations for a strain value corresponding to the 11 ± 2 nm Si shell on 15 nm Ge core. The intensities are normalized to one, and the FWHM was taken as the experimental FWHM. In the calculated spectrum in the $\bar{x}$(y,y)x configuration, both the TO and LO peaks are visible, however, in our experiments we see only one broader downshifted peak. This is likely due to the dielectric mismatch between the NWs and their surrounding environment and/or to a relaxation of the Raman selection rules because of the size and geometry of the probes sample.[55,56] Indeed, the broader linewidths in the $\bar{x}$(y,y)x configuration and the limited spectral resolution of our setup may further hinder the detection of closely spaced modes. We performed the polarization resolved measurements on CS NWs with different Si shell thicknesses and plotted the difference $\omega_{LO}-\omega_{TO}$ as a function of Si shell thickness (see Figure 5 (c)). The measurements are repeated on multiple individual wires of the same sample design, and the error bars represent the standard deviation. The corresponding theoretical splitting as a function of Si shell thickness is shown in Figure 5(d). Both experiment and theory reveal the same qualitative and quantitative trend: the splitting increases with thicker Si shells, i.e., with increasing compressive strain.

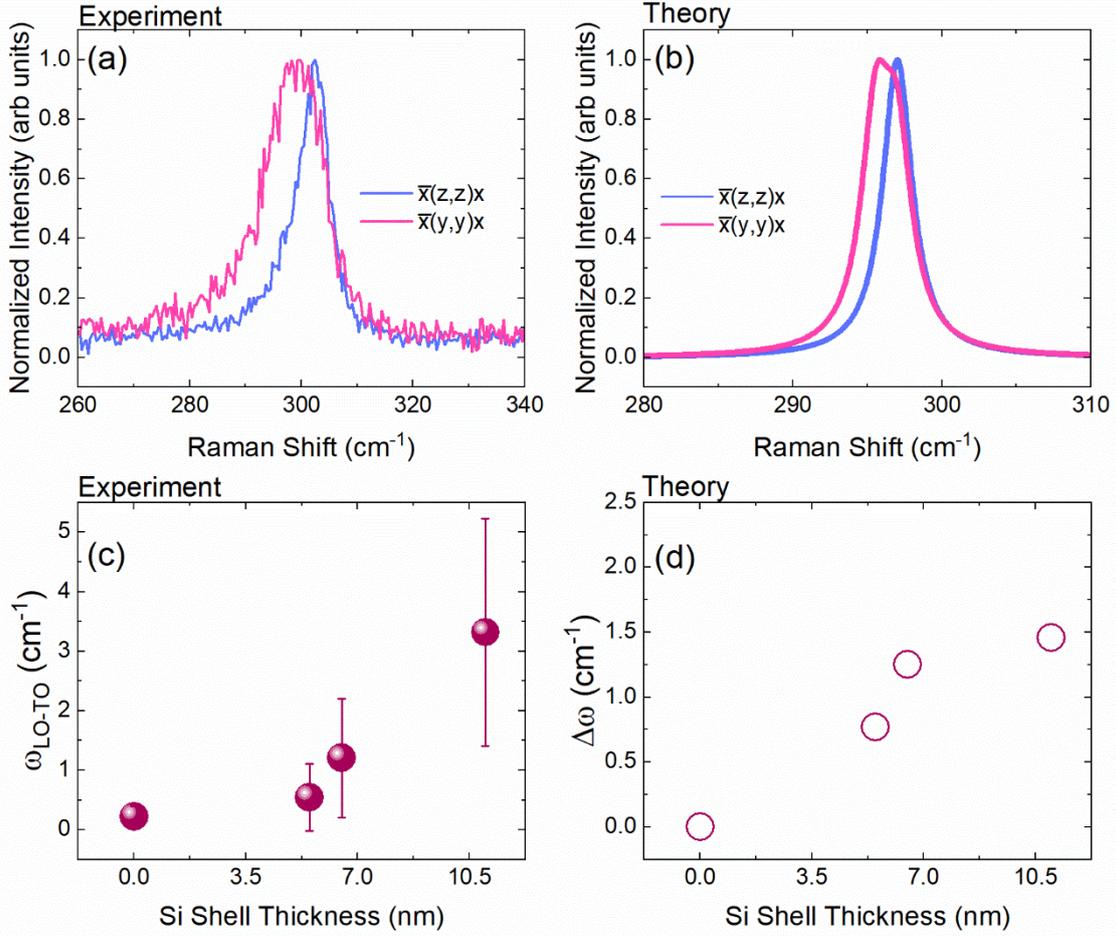

**Figure 5.** (a) Experimental polarized Raman spectra ($\bar{x}(z,z)x$ and $\bar{x}(y,y)x$ configurations) of Ge/Si CS NWs showing mode splitting due to anisotropic strain. This exemplary measurement was done on CS NW with 15 nm Ge core and 11 ± 2 nm Si shell. (b) Corresponding theoretical Raman spectra for the same polarization configurations in strained bulk Ge with the same strain level as measured in CS NW with 15 nm Ge core and 11 ± 2 nm Si shell, showing good qualitative agreement with the experimental peak positions and asymmetry. (c) Experimentally extracted LO/TO phonon splitting ($\omega_{LO}-\omega_{TO}$) as a function of Si shell thickness, indicating increasing strain-induced splitting with shell thickness. (d) Theoretically predicted frequency difference between $\bar{x}(z,z)x$ and $\bar{x}(y,y)x$ polarized Raman modes as a function of shell thickness, showing a similar trend to the experimental data.

2.3. Transport Measurements:

**Figure 6**(a) presents a three-dimensional schematic of a representative device used for mobility measurements. The devices are fabricated on a highly boron-doped Si substrate (light grey), which is covered by a 300 nm thick silicon oxide layer (dark grey), enabling the substrate to function as a global back gate. The NW core and shell are indicated in red and blue, respectively, while the Ti/Pd ohmic contacts are shown in green. A DC source–drain bias voltage $V_{SD}$ is applied to the source while the drain is grounded, and the resulting transport current $I_{SD}$ through the NW is measured. The inset of Figure 6(a) shows a zoomed-in SEM image of a single NW device, where both ends of the NW are covered by metallic ohmic contacts. In total, 22 NWs were characterized, with channel lengths ranging from 300 to 500 nm and NW diameters between 20 and 50 nm.

The transport current $I_{SD}$ was measured at 4.2K as a function of the back gate voltage $V_G$, to obtain the corresponding IV curves. The NW quality is then assessed by extracting the field-effect mobility using the expression applicable to the diffusive transport regime[57]:

$$G(V_G) = \left(R_C + \frac{L^2}{\mu_h C(V_G - V_{th})}\right)^{-1} \qquad (4)$$

where, $\mu_h$ is the hole mobility and $L$ is the channel length, $G(V_G) = I_{SD}/V_{SD}$ is the measured conductance of the NW, $V_{th}$ is the pinch off voltage, $V_G$ is the applied back gate voltage, and $R_C$ is the contact resistance. The channel capacitance $C$ was determined using COMSOL simulations, as detailed in the Supporting Information. By fitting this equation to the measured conductance, we extracted the carrier mobility, which is determined by the slope of the curve near pinch-off, and the contact resistance, which is determined by the saturation value of the conductance. This was performed individually for several bias voltages and subsequently averaged to obtain the final mobility and contact resistance for each NW.

Figures 6(b) and 6(c) show two examples of the dependence of the source–drain current $I_{SD}$ on the back gate voltage $V_G$ for NWs with a 5 nm shell thickness and diameters of 30 nm and 20 nm, respectively, measured at several applied source–drain bias voltages. The experimental data are accompanied by fits to Eq. 1. Deviations from ideal transport behaviour are attributed to defect scattering within the NWs, resulting in the spread reflected by the error bars in Figure 6(d). As a representative device, the 30 nm-diameter NW exhibits a carrier mobility of approximately 9,400 cm² V⁻¹ s⁻¹, whereas the 20 nm-diameter NW device exhibits a record-high carrier mobility of 25,400 cm² V⁻¹ s⁻¹.

All NWs measured belonged to the same growth run, the results of which are plotted as a function of NW diameter in Figure 6(d). The main finding here is the significantly higher value for both maximum and average transport mobilities, compared to previous reports.[58–61] In addition, we confirm earlier reports[58] that thinner NWs tend to exhibit higher mobilities, as seen in Figure 6(d).

This in principle provides a practical guideline for NW selection under an optical microscope. In addition, for both one-dimensional and three-dimensional transport in the NWs, the extracted mean free path ranges from approximately 12 nm to 270 nm, further details are provided in the supporting information.

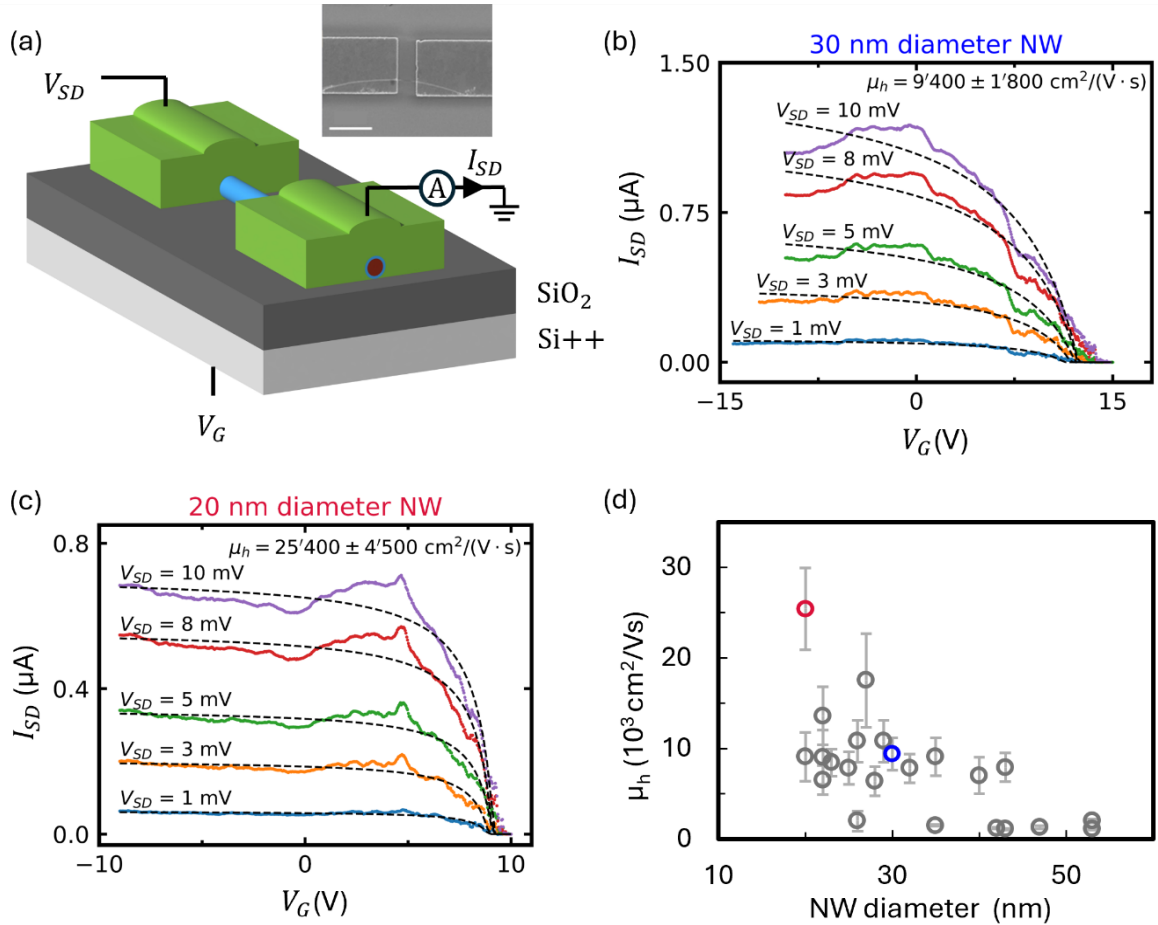

**Figure 6**. (a) Device schematic used for mobility measurements. The device is fabricated on a highly boron-doped Si substrate (light grey) covered by 300 nm of silicon oxide (dark grey), serving as a global back gate. The NW core and shell are indicated in crimson and blue, respectively, and the Ti/Pd ohmic contacts are shown in green. A DC source–drain bias voltage $V_{SD}$ is applied, and the resulting transport current $I_{SD}$ through the NW is measured. The inset shows a zoomed-in scanning electron microscopy (SEM) image of a single NW device with both ends contacted by metallic ohmic electrodes. The scale bar is 1 $\mu m$. Current ($I_{SD}$) versus back gate voltage ($V_G$) of a 30 nm (b) and 20 nm (c) with a 5 nm shell for various source–drain bias voltages, as labelled. Error bars on the mobility of up to 20% are dominated by the large (repeatable) current fluctuations. (d) Overview of all the extracted mobilities, $\mu_h$

as a function of diameter, highlighting the highest mobility (red circle, 20 nm diameter) and a typical mid-range mobility (blue circle, 30 nm diameter).

3. Conclusion

In this work we performed a systematic and comprehensive characterization of the strain in Ge/Si CS NWs through Raman spectroscopy and GPA analysis of HRTEM micrographs. Raman spectroscopy was performed on individual Ge/Si CS NWs with statistically significant sampling. The compressive strain in the Ge core was quantitatively extracted from the Raman peak shifts. The results reveal that the compressive strain in the Ge core increases with increasing Si shell thickness. For a fixed Si shell thickness, varying the Ge core diameter from 5 nm to 20 nm did not result in a significant change in the observed strain. In addition, the LO and TO phonon mode splitting was clearly resolved using polarization resolved Raman measurements and quantitatively analysed. The LO-TO splitting was further corroborated through ab initio calculations, providing theoretical support for the experimental observations and confirming the dependence of strain on shell thickness. GPA reveals tensile strain in the Si shell with respect to the Ge core, while Raman spectroscopy indicates compressive strain in the Ge core itself. It is worth noting, however, that GPA measures relative strain, and in this study, the Ge core was taken as the reference relaxed lattice for the GPA analysis of the Si shell. In contrast, Raman spectroscopy gives access to the absolute strain in the measured material, in this case, the Ge in the NW core. Together, these complementary measurements confirm coherent strain transfer across the core and shell. Furthermore, from the Raman measurements on our CS NWs with 11 ± 2 nm Si shell, we observed significant Fano broadening, comparable to that in doped – i-Ge/p-Si CS NWs with hole densities up to ~$10^{18}$ cm$^{−3}$. Finally, we presented the hole field effect mobility in Ge/Si CS NW electronic devices and found a clear dependence on the NW diameter. The average mobility 8,000 cm² V$^{−1}$ s$^{−1}$ over 22 NWs was improved compared to the previously reported peak mobilities[58–61], and we have obtained an individual NW with record high hole mobility of up to 25,400 cm² V$^{−1}$ s$^{−1}$. These results suggest that our CS NWs are of very good crystalline quality and structural integrity, with a low density of impurities and defects.

4. Experimental Section/Methods

*CVD Growth of CS NWs*: The preparation of the substrate consisted of two steps. First, commercial Au nanoparticles (SPI Supplies and BBI Solutions) with a diameter ranging from 5 to 20 nm were deposited on a Si <100> wafer (float zone, undoped, resistivity >10,000

Ohm·cm, University Wafer Inc., South Boston, MA, USA). The deposition was triggered by an electrostatic approach upon the addition of HCl 0.1 M.[62] Second, the native silicon oxide was removed by fully immersing the samples into a 2.3% HF aqueous solution for 1 min. The substrates were then loaded into the load lock of the growth chamber within less than 10 min. Subsequently, the NWs were grown in a PlasmaPro 100 Nanofab reactor from Oxford Instruments, Wiesbaden, Germany (base pressure <0.5mTorr). The commercially available precursor gases employed were germane gas ($GeH_4$, PanGas AG, Switzerland, 99.999%) and silane gas ($SiH_4$, PanGas AG, Dagmersellen, Switzerland, 99.999%), while the carrier gases we used were argon and hydrogen (Ar, $H_2$, PanGas AG, Switzerland, 99.999%).

The growth protocol for the Ge NWs consisted of six steps: (i) preheating the chamber; (ii) loading the sample immediately after HF; (iii) heating the substrate to 360°C in $H_2$ environment (total flow 100 sccm) at a pressure of 0.5 Torr for 3 min; (iv) addition of the reaction gas mixture consisting of 10% $GeH_4$ in Ar (total flow 50 sccm) at a total pressure of 2 Torr for 15 min; (v) temperature decrease to 280°C with a rate of 2.4 °C/min, keeping the gas flow and pressure constant; (vi) continuation of the growth for 86 min at 280°C, leading to 120 min starting with step (v).

The Ge-Si CSNW were grown instead according to a seven step process: (i) preheating the chamber; (ii) loading the sample immediately after HF; (iii) heating the substrate to 350°C in $H_2$ environment (total flow 100 sccm) at a pressure of 0.5 Torr for 3 min; (iv) addition of the reaction gas mixture consisting of 10% $GeH_4$ in Ar (total flow 200 sccm) at a total pressure of 2 Torr for 15 min; (v) pressure increase to 10 Torr keeping the flow constant for 1 min; (iv), reduction of the total flow to 70 sccm, keeping the temperature and pressure constant for 45 min; (v) interruption of the reaction gas mixture; (vi) cool down the chamber to 250°C in $H_2$ environment (total flow 100 sccm) for 25 min; (vii) introduction of the reaction gas mixture of 2% $SiH_4$ in $H_2$ (total flow 204 sccm) at a total pressure of 0.5 Torr, employing a plasma with a power of 22 W.

*TEM and EDX*: Electron transparent cross-sections of the NW were obtained using $Ga^+$ ions in a ZEISS Crossbeam 540 FIB/SEM operated at 30, 5 and 2 kV, using beam currents ranging from 3 nA to 10 pA. Final polishing was performed using 10 pA at 2 kV. A 20 nm thick Carbon layer was evaporated onto the chip containing the NWs to minimize charging effects, before 100 nm of protective Pt was deposited by focused electron beam induced deposition, followed by a thicker (~1.5 µm) Pt layer deposited using focused ion beam induced deposition. HR-TEM and STEM-EDX measurements were subsequently performed on a JEOL JEM F200 cFEG operated at 200 kV and equipped with a EMSIS XAROSA CMOS camera and the JEOL JED-2300 analysis station for EDS with 5 nm resolution.

*GPA:* Geometric phase analysis (GPA) was calculated by using the FRWR tools plugin developed for Gatan Digital Micrograph by Christoph Koch at the Humboldt University zu Berlin and subsequently plotted using custom Python scripts. The Ge core of each NW was selected as zero-strain reference in each case. The quantitative relative strains were obtained from cropped TEM images, after rotating them to align the axial and transverse NW crystal axis with the image frame.

*μ-Raman Spectroscopy*: We used a 633 nm laser to excite the sample and a 100x objective with a high numerical aperture (0.95) for focusing the excitation laser and collecting the scattered light. The Raman measurements were performed in backscattering geometry with the help of a Horiba T64000 triple spectrometer in subtractive mode with a 1.800 g/mm grating and a liquid-nitrogen-cooled CCD detector. We used an excitation laser power below 10 μW to perform the Raman experiments.

*DFPT calculations:* We performed density-functional calculations with the ABINIT code[63] to obtain the ground-state geometry and the susceptibility Raman tensors. We used a plane wave cutoff of 41 Ha, a grid to sample the Brillouin zone of 16x16x8 **k**-points, and the Local Density Approximation (LDA) for the exchange-correlation functional. We used the non-primitive 6-atom Ge unit cell, where the *z*-axis is parallel to the [111] crystal axis (i.e., the growth direction of the CS NWs). To simulate the compressive strain caused by the Si cell, we reduced the **c**-vector (leaving the **a**- and **b**-vector frozen) and reoptimized the atomic positions.

*Mobility Device Fabrication*: A conductive boron-doped Si substrate was used for the fabrication of the mobility devices, serving simultaneously as the back gate. Alignment marks and bonding pads were first patterned on the substrate, which was subsequently diced into 5 mm × 5 mm small substrates. NWs were then transferred onto the entire substrate surface and baked at 185 °C for 5min to attach the NWs the chip. A layer of adhesion promoter (AR300-80) was spin-coated at 4000 rpm (acceleration: 1000 rpm s$^{-1}$) for 40 s and baked at 185 °C for 120 s. The residual adhesion promoter was removed by immersing the device in acetone for 7 min followed by isopropanol (IPA) for 1 min.

Subsequently, two layers of EL6 resist were spin-coated under identical conditions (4000 rpm, 1000 rpm s$^{-1}$ acceleration, 40 s duration) and baked at 185 °C for 90 s. Optical microscopy was employed to image the NWs and identify those with a [110] crystal orientation for device fabrication. [110]-orient NWs are believed to accommodate stain without nucleating defects [58]. The ohmic contact pattern was designed based on the selected NWs, and electron-beam lithography (EBL) was used to write the ohmic contacts. After EBL, the resist was developed in AR500-60 for 60 s and rinsed in IPA for 10 s. A brief plasma ashing step (30 W, 10 s) was performed to remove the residual, followed by immersion in a 10:1 buffered Hydrofluoric acid (BOE 10:1) to remove the native $SiO_2$ layer on the NW surface. Titanium (0.3 nm) and Palladium (50 nm) layers were then deposited by electron-beam evaporation to form the ohmic contacts. The resistance of individual NWs was measured using a needle probe to identify functional NWs suitable for subsequent electrical measurements. Conductive silver paint was used to mount the fabricated device onto a *Kyocera* sample holder. Additional silver paint was applied along the sides of the device to ensure electrical contact between the doped Si substrate and the bottom metal plate of the sample holder. Finally, the bond pads of every functional NW and the bottom metal plate were wire-bonded to the corresponding pins on the holder to complete the electrical connections and enable back gate tunability.

*Mobility Measurement:* The device was mounted on a sample holder and cooled in liquid helium to a temperature of 4.2 K. Measurements were performed by grounding one ohmic contact of the NW and applying a source–drain bias ($V_{SD}$) through a current–voltage converter (IVC, *BASPI SP983c*), while measuring the transport current ($I_{SD}$) through the NW by digital multiplier (DMM, *Agilent 34410A*). Subsequently, a voltage was applied via Digital-to-Analog Converter (DAC, *BASPI SP927*) to the back gate ($V_G$) to measure the carrier density. By measuring the transport current as a function of back gate voltage for different source–drain biases, we obtained current-voltage (IV) characteristics of every NW. Fitting these curves within the transport model [57] the field-effect mobility and contact resistance for individual NWs were extracted.

**Acknowledgements**

This work was supported as a part of NCCR SPIN, a National Centre of Competence in Research, funded by the Swiss National Science Foundation (grant number 225153). R.R. acknowledge financial support by MCIN/AEI/10.13039/501100011033 under grant PDC2023-145934-I00, and the Severo Ochoa Centres of Excellence Program under grant CEX2023-001263-S, and by the Generalitat de Catalunya under grant 2021 SGR 01519. We thank the Centro de Supercomputación de Galicia (CESGA) for the use of their computational resources.  A.K.S. acknowledges financial support from the Georg H. Endress foundation and the Nachwuchsförderungs der Universität Basel. We acknowledge EU's H2020 Marie

Skłodowska-Curie Actions (MSCA) cofund Quantum Science and Technologies at the European Campus (QUSTEC) grant no. 847471, and the UpQuantVal InterReg.

We thank the Swiss Nanoscience Institute (SNI) for providing the Nano Imaging facilities.

We thank Dr. Cedric Gonzales for his help in sample preparation for Raman measurements.

**Data Availability Statement**

All data within the article and the Supplementary Information that support the findings of this study are openly available in ZENODO at (*to be added upon acceptance*), Reference No. (*to be added upon acceptance*).

**Supporting Information**

Supporting Information is available from the Wiley Online Library or from the author.

# Supporting Information

Quantifying Strain and its Effect on Charge Transport in Ge/Si Core/Shell Nanowires

*Aswathi K. Sivan, Nicolas Forrer, Aakash Shandilya, Yang Liu, Alexander Vogel, Arianna Nigro, Janica Böhler, Pierre Chevalier Kwon, Artemii Efimov, Ilya Golokolenov, Gerard Gadea, Riccardo Rurali, Andreas Baumgartner, Dominik Zumbühl, and Ilaria Zardo*[*]

**Section S1: Structural characterization of the CS NWs**

In this section we present structural characterization for two additional CS NWs, more specifically for shell thicknesses of 5.5 ± 1 nm and 15.1 ± 3 nm. In Figure S1 (a) HR-TEM is presented, indicating the high crystal quality of the samples, whereas for the 15 nm thick shell (b) a poly-crystalline growth is observed, resulting from the relaxation of the strain due to the lattice mismatch of the two materials. EDX line scans (c) highlight the presence of the target materials Si and Ge distributed in the shell and the core, respectively. Please note that panel (b) shows a top-down view instead of a cross-sectional one.

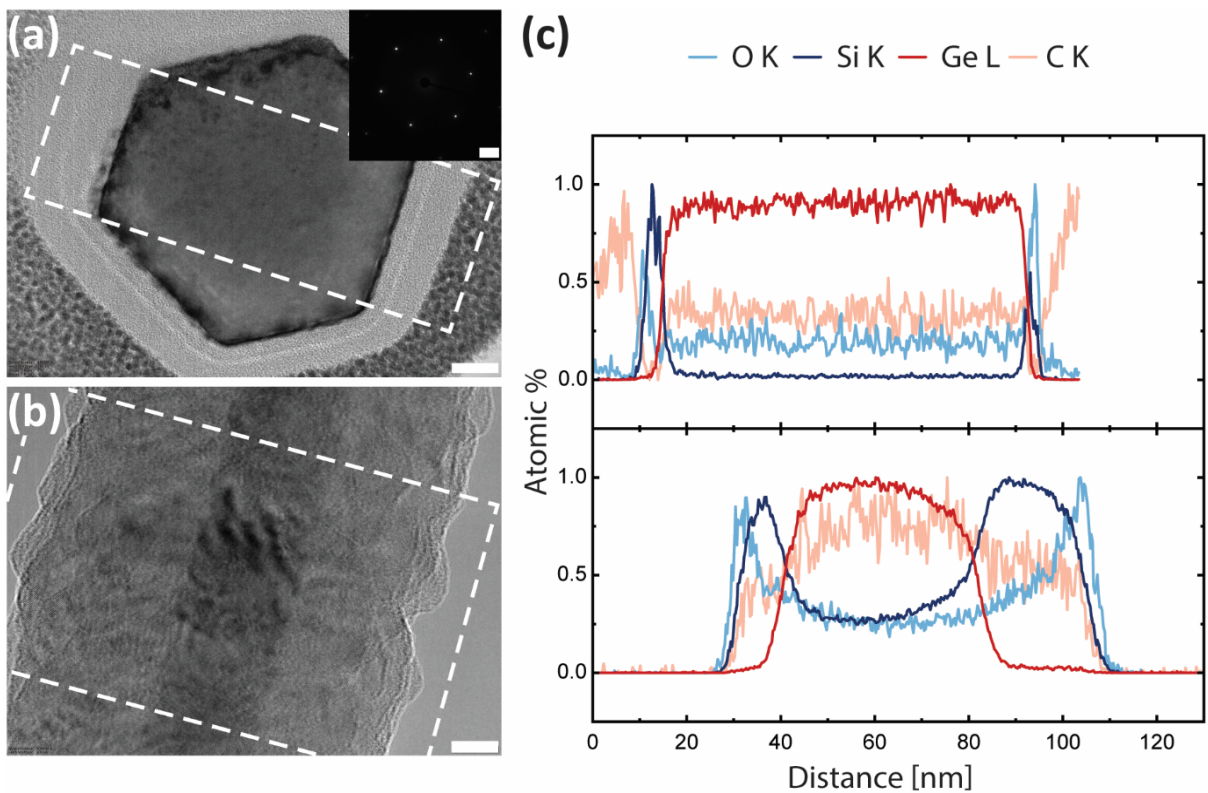

Figure S 1. (a) Cross-sectional and (b) top-down HR-TEM images of Ge/Si CS NWs grown from a 15 nm Au colloid with shell thickness of 5.5 ± 1 nm and 15.1 ± 3 nm, respectively. The inset in (a) shows a SAED pattern obtained on the cross-section. The scale bar for the HR-TEM

images and for the SAED patterns is 10 nm and 2 nm$^{-1}$ in each panel, respectively. (c) EDX line profiles of the O-K, Si-K, Ge-L, and C-K edges obtained on the HR-TEM images, as indicated in panels (a) – (b).

**Section S2. µ-Raman Spectroscopy:**
**Statistics of the Measurements:**

Figure S2 shows the statistics of the CS NWs with different Si shell thicknesses. To quantify wire-to-wire variability, we measured the Ge–Ge Raman peak for Ge/Si core–shell NWs with Si-shell thicknesses of 5.5 ± 1 nm, 6.5± 2 nm, and 11 ± 2 nm. For each set of Si shell thickness, the NWs were transferred from the as-grown sample onto a TEM grid prior to Raman on individual NW measurement. These grids were then used for microscopy analysis. For each condition, individual peak positions cluster around the batch mean (dotted line), with the grey band indicating ± SD.

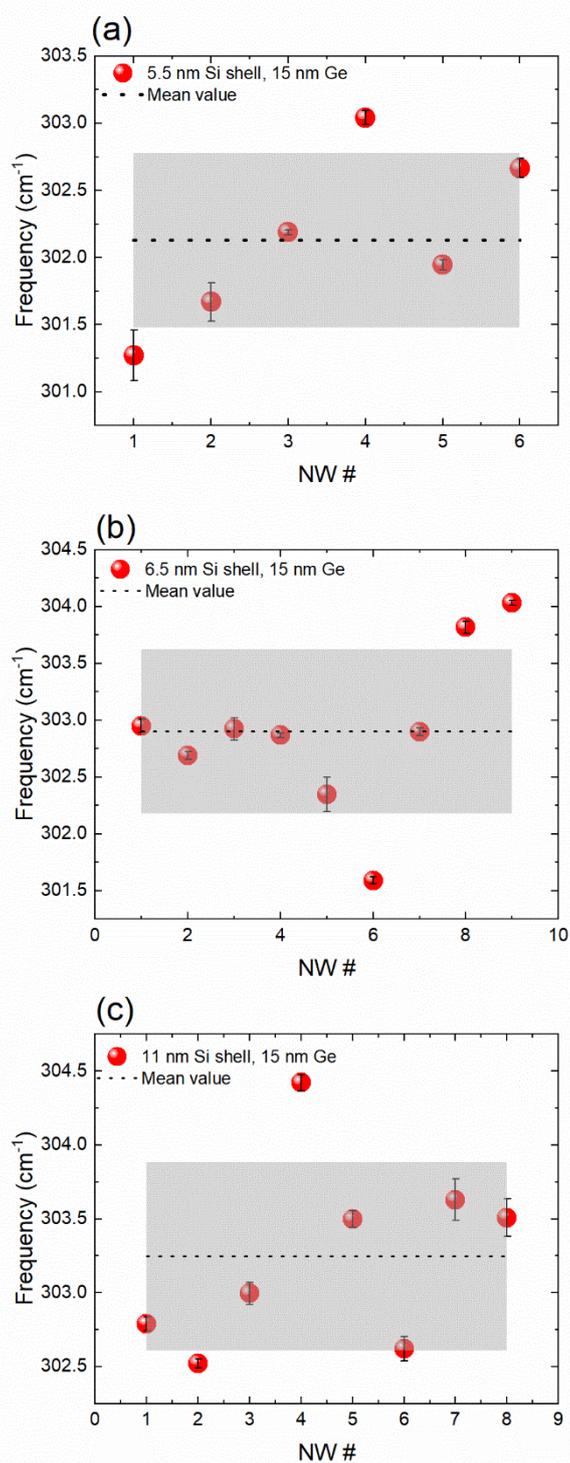

S 2. Wire-to-wire statistics of the Ge–Ge Raman peak for Ge/Si CS NWs with different Si-shell growth times. (a) 5.5 nm Si shell, (b) 6.5 nm Si shell, and (c) 11 nm Si shell; Ge cores grown using 15-nm Au colloids. Red circles are peak positions from individual NWs (error bars from the line-shape fit); the black dotted line marks the mean peak position from each sample, and the grey band shows ± standard deviation. All wires in a panel were taken from the same as-grown sample and measured under identical conditions.

**Raman measurements on a 15.1 nm thick Si shell growth on 15 nm Ge core:**

Extending the Si-shell growth to 15 mins created a 15.1 ± 3 nm thick Si shell and produces a red shift of the Ge–Ge Raman peak, consistent with defect-mediated relaxation of the compressive core strain.

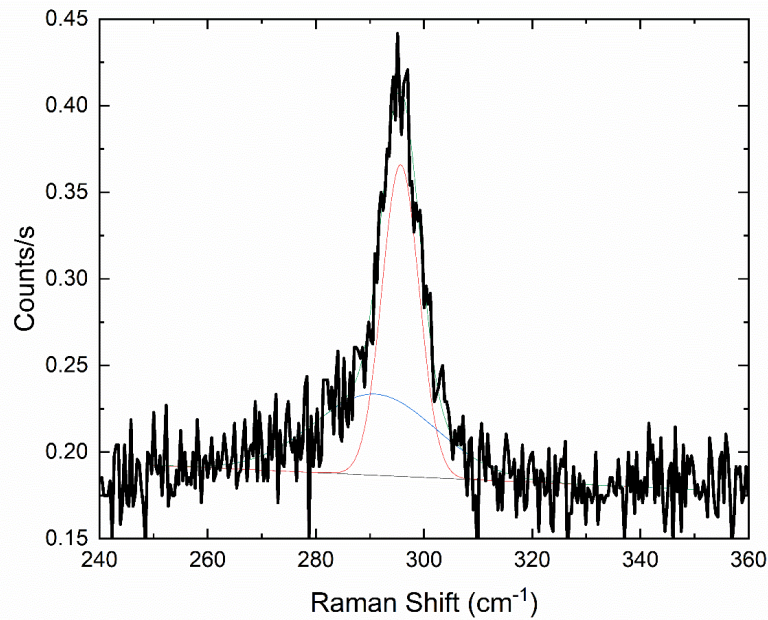

S 3 The Raman measurement on a 15.1 ± 3 nm Si shell CS NW, here the Raman peak is red shifted maybe attributed to the onset of defect-driven relaxation in 15.1 ± 3 nm Si shells.

**Choice of the fit – Fano vs Lorentz function**

In this work, we have used a Lorentz function to fit the Ge-Ge phonon mode for the bare Ge-NWs and a Fano function for the Ge-Ge phonon mode for the Ge/Si CS NWs. The peak is weakly asymmetric, with a longer low-frequency tail. The Fano lineshape accounts for the interference between the discrete phonon and the hole continuum.

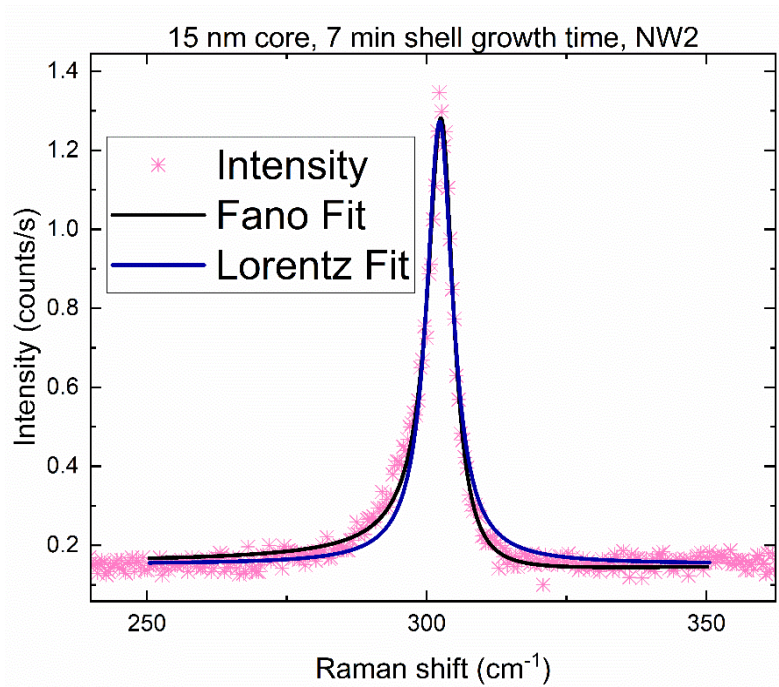

Figure S 4. Fano (black) reproduces the slight asymmetry and low-frequency tail of the Ge peak; the Lorentzian (blue) underestimates intensity on the low-frequency side.

**Section S3. Mobility measurements:**

**Simulation of capacitance between NWs and back gate**

As mentioned earlier, both ends of each NW are connected by Ti/Pd ohmic contacts. Since the NW itself is conductive, it, together with its contacts, effectively forms a dumbbell-shaped metallic structure above the gate. The P-doped Si substrate is also conductive and serves as a global back gate. To evaluate the gate–NW coupling, we perform COMSOL simulations of the capacitance between this metallic structure and the back gate, modelling the NW and its contacts as a single electrode for computational simplicity. The geometric parameters—including the NW length between contacts (L), contact width (W), and NW diameter (D)—are extracted from SEM images. Based on TEM and EDX analyses, the Ge core radius is taken as two-thirds of the total NW radius, with the remaining one-third corresponding to the Si shell. The ohmic contact length is set to 4 μm, sufficiently long compared with the active NW segment, and the $SiO_2$ thickness on the substrate is fixed at 300 nm. For each device, these geometric parameters are determined individually, and the model is then used to simulate the corresponding capacitance prior to extracting the mobility. Figure S5 shows a schematic top view and cross-section of the mobility device used in the simulation: the dark blue region corresponds to the P-doped conductive substrate, the light blue layer represents the 300 nm $SiO_2$ on the substrate, and the grey boxes indicate the Ti/Pd ohmic contacts. In the NW, the orange region denotes the Ge core, while the green region corresponds to the Si shell.

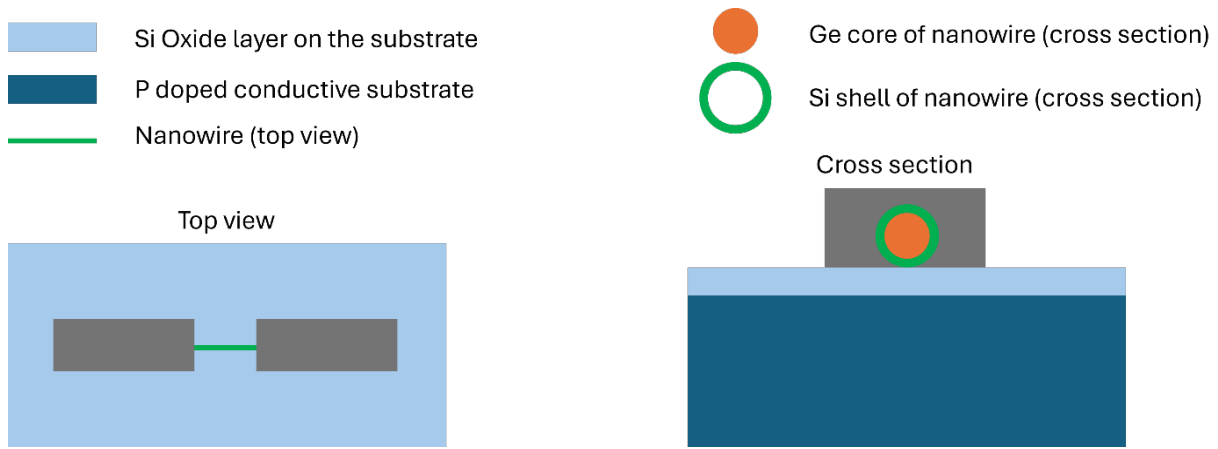

Figure S 5. Schematic top view and cross-section of the mobility device used in the simulation. The light grey region represents the P-doped conductive substrate, while the dark grey layer corresponds to the 300 nm SiO₂ on the substrate. The green boxes indicate the Ti/Pd ohmic contacts. For the NW, the crimson region denotes the Ge core, and the blue region denotes the Si shell

**Details of NW Parameters**

| Sr. No. | Length (nm) | Diameter (nm) | Mobility (cm2/Vs) | Pinch-off voltage (V) | Contact resistance (kΩ) | Total resistance ($V_G = 0$) (kΩ) | Mean free path 1V above depletion (nm) |
|---|---|---|---|---|---|---|---|
| 1 | 1,280 | 40 | 7,000 ± 2,000 | 4.6 ± 0.9 | 18.2 ± 1.5 | 38 | 126 ± 36 |
| 2 | 505 | 53 | 2,000 ± 200 | 6.2 ± 0.3 | 16.2 ± 1.3 | 50 | 23 ± 2 |
| 3 | 890 | 35 | 1,500 ± 100 | 8.2 ± 0.5 | 21.1 ± 0.4 | 33 | 23 ± 1 |
| 4 | 510 | 43 | 7,900 ± 1,600 | 7.1 ± 0.2 | 7.2 ± 0.3 | 11 | 100 ± 20 |
| 5 | 375 | 43 | 1,100 ± 100 | 8.2 ± 0.6 | 19.7 ± 1.4 | 50 | 12 ± 1 |
| 6 | 1,030 | 32 | 7,800 ± 1,600 | 7.5 ± 0.2 | 12 ± 1 | 18 | 128 ± 26 |
| 7 | 590 | 53 | 1,100 ± 100 | 7.1 ± 0.4 | 16.9 ± 1.2 | 50 | 17 ± 1 |
| 8 | 1,570 | 47 | 1,300 ± 200 | 0.2 ± 0.3 | 57 ± 2 | 5,000 | 26 ± 4 |
| 9 | 640 | 30 | 9,400 ± 1,800 | 11.9 ± 0.5 | 7.2 ± 0.7 | 8 | 127 ± 24 |
| 10 | 370 | 23 | 8,400 ± 1,500 | 14.2 ± 0.2 | 5.4 ± 0.6 | 6 | 80 ± 14 |

| 11 | 550 | 27 | 17,500 ± 5,200 | 11.1 ± 0.2 | 8.3 ± 0.8 | 9 | 210 ± 62 |
|---|---|---|---|---|---|---|---|
| 12 | 480 | 22 | 9,000 ± 3,000 | 13.4 ± 0.1 | 10 ± 0.8 | 12 | 100 ± 33 |
| 13 | 500 | 25 | 7,800 ± 1,800 | 12.6 ± 0.3 | 7.2 ± 0.6 | 9 | 90 ± 20 |
| 14 | 660 | 22 | 6,500 ± 1,600 | 7.5 ± 0.5 | 26 ± 1 | 32 | 82 ± 20 |
| 15 | 420 | 22 | 13,600 ± 3,200 | 9 ± 0.1 | 17 ± 2 | 18 | 133 ± 31 |
| 16 | 530 | 20 | 25,400 ± 4,500 | 9 ± 0.1 | 14.5 ± 1.5 | 18 | 275 ± 49 |
| 17 | 300 | 35 | 9,100 ± 2,100 | 4.8 ± 0.2 | 6.7 ± 0.6 | 12 | 80 ± 18 |
| 18 | 450 | 26 | 2,000 ± 1,100 | 1.2 ± 0.8 | 25 ± 3 | 184 | 21 ± 11 |
| 19 | 500 | 26 | 10,800 ± 2,300 | 1.8 ± 0.2 | 10 ± 1 | 39 | 120 ± 25 |
| 20 | 400 | 20 | 9,100 ± 2,700 | 5.1 ± 0.3 | 6.5 ± 0.2 | 13 | 79 ± 23 |
| 21 | 550 | 28 | 6,400 ± 1,600 | 3.3 ± 0.2 | 14 ± 1 | 26 | 69 ± 17 |
| 22 | 550 | 29 | 10,800 ± 2,300 | 5.2 ± 0.2 | 5.7 ± 0.5 | 12 | 118 ± 25 |

Table S 1. Detailed parameters of the measured NWs

Table summarizes the parameters of all 22 NWs measured in this work. The NW length is defined as the distance between the two ohmic contacts, which was extracted from SEM images together with the diameter. The total resistance is measured at $V_G = 0$. The NW–back gate capacitance was simulated using COMSOL, which was used to extract the pinch-off voltage, contact resistance, and carrier mobility by fitting to Eq. 1. The mean free path is estimated as detailed below at a gate voltage of 1 V before depletion.

**Mean free path**

Using a simple Drude-Sommerfeld model, the mean free path for one-dimensional (1D) transport is given

$$l = \mu_h n \frac{\pi \hbar}{2e},$$

while for three-dimensional (3D) transport is

$$l = \mu_h \sqrt[3]{n} \frac{\hbar \sqrt[3]{3\pi^2}}{e}.$$

Here, $n$ is the charge carrier density, which depends on the gate voltage according to

$$n = \frac{c(V_G - V_{th})}{e},$$

where $c$ is the channel capacitance per unit length for the 1D case and per unit volume for the 3D case. Using the parameters of our NWs, we find that the mean free paths obtained from the 1D and 3D models are quite similar and in good agreement. The mean free path calculated using the 1D expression is listed in Table S1 and plotted as a function of mobility in Figure S6 at a gate voltage of 1 V before pinch-off using a simple Drude model. Notably, farther away from pinch-off, the extracted mean free path can exceed the channel length. In principle, this could imply the emergence of quantized conductance which, however, is not observed in the measurements.

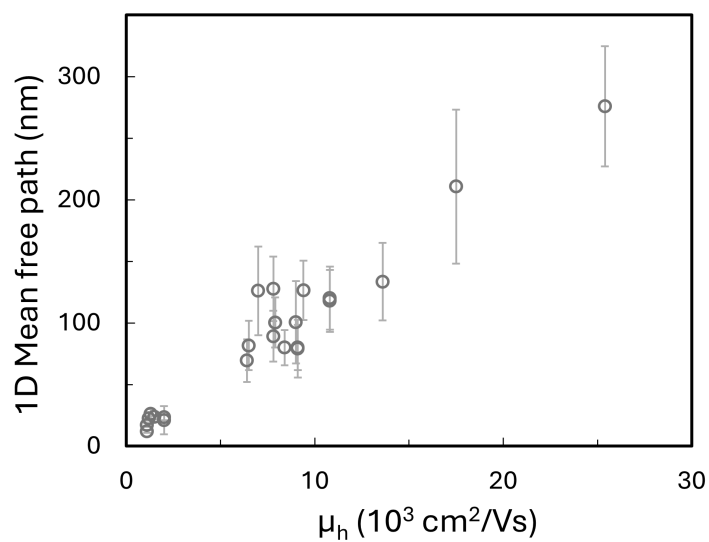

Figure S 6. 1D mean free path of the 22 NWs as a function of carrier mobility. The mean free path values are obtained using the parameters of our NWs and are listed in Table S1. The results show good agreement with those derived from the three-dimensional (3D) model.